\documentclass[prl,aps,reprint,showpacs,superscriptaddress]{revtex4-1}

\usepackage{amsmath}
\usepackage{amsfonts}

\usepackage[pdftex]{graphicx}
\usepackage[pdftex]{epsfig}
\usepackage{epstopdf}
\usepackage{dcolumn}
\usepackage{bm}
\usepackage{color}

\newcommand{\beq}{\begin{equation}}
\newcommand{\eeq}{\end{equation}}
\newcommand{\beqa}{\begin{eqnarray}}
\newcommand{\eeqa}{\end{eqnarray}}
\newcommand{\vep}{\varepsilon}
\newcommand{\rd}{|\negthickspace \Downarrow \rangle}
\newcommand{\ld}{\langle \Downarrow \negthickspace |}
\newcommand{\ru}{|\negthickspace \Uparrow \rangle}
\newcommand{\lu}{\langle \Uparrow \negthickspace |}

\begin{document}

\title{Controllable anisotropic exchange coupling between spin qubits in quantum dots}
 \author{Yun-Pil Shim}
 \affiliation{Department of Physics, University of Wisconsin-Madison, Madison WI 53706, USA}
 \author{Sangchul Oh}
 \affiliation{Department of Physics, University at Buffalo, State University of New York, Buffalo, NY 14260, USA}
 \author{Xuedong Hu}
 \affiliation{Department of Physics, University at Buffalo, State University of New York, Buffalo, NY 14260, USA}
 \author{Mark Friesen}
 \affiliation{Department of Physics, University of Wisconsin-Madison, Madison WI 53706, USA}
 \date{\today}

\begin{abstract}
The exchange coupling between quantum dot spin qubits is isotropic, which restricts the types of quantum gates that can be formed.
Here, we propose a method for controlling anisotropic interactions between spins arranged in a bus geometry.
The symmetry is broken by an external magnetic field, resulting in XXZ-type interactions that can efficiently
generate maximally entangled Greenberger-Horne-Zeilinger (GHZ) states or universal gate sets for exchange-only quantum computing.
We exploit the XXZ couplings to propose a qubit scheme, based on double dots.
\end{abstract}

\pacs{03.67.Lx, 73.21.La, 75.10.Pq, 03.67.Hk, 03.67.Bg}
\maketitle

Electron spins in quantum dots are among the most promising candidates for qubits, due to their perceived scalability and integrability with
current semiconductor technologies \cite{Loss1998}, and their long coherence and relaxation times \cite{Amasha2008,Morello2010,Bluhm2010}. The
spin-dependent component of the Coulomb interaction, known as exchange coupling, has arisen as the most prominent tool for coupling spin qubits,
since it is fast and controllable via electrostatic top-gates \cite{Petta2005}. The exchange interaction produces an isotropic Heisenberg
coupling between the spins, which by itself does not provide a universal set of quantum gates, due to its high symmetry
\cite{Bacon00,divincenzo_bacon_nature2000}. The existence of a lower-symmetry coupling would therefore enhance the toolbox for spin qubits, while
providing new opportunities for efficient gating. For example, we consider the exchange-only gating scheme of DiVincenzo \textit{et al.}, which
encodes three physical qubits into a logical qubit \cite{divincenzo_bacon_nature2000,Kempe01}. If the Heisenberg symmetry is reduced from XXX
(isotropic) to XXZ (axial), then the gating requirements will be ameliorated, both in terms of the number of physical qubits and the number of
gate operations \cite{lidar_wu_prl2001}. The XXZ interaction also provides added entangling power, since it can be used to
efficiently generate a multi-qubit, maximally entangled Greenberger-Horne-Zeilinger (GHZ) state
\cite{galiautdinov_martinis_pra2008}; this does not appear to be possible with isotropic interactions.
Spin-orbit coupling also produces a form of anisotropic interaction between spins in semiconductor quantum dots \cite{Kavokin_PRB01},
which can be used to generate universal quantum gates \cite{Stepanenko_PRL04}.
While the effective interaction depends on magnetic field, it is difficult to tune \cite{Chutia},
and in materials like silicon, its effect is almost negligible.

In this paper, we demonstrate how a controllable anisotropic exchange interaction can be realized for electron spins in semiconductor
nanostructures.  More specifically, we show that XXZ couplings can be obtained and controlled in different quantum dot geometries.
The full range of new behaviors we explore can be observed in a spin bus geometry,
which is a linear array of quantum dots with strong, static interactions, and weakly coupled external qubits.
Such architectures can potentially overcome the severe short-range
nature of the exchange coupling and implement efficient multi-qubit entanglement.
Here, we are particularly interested in adiabatic gate operations involving the bus
ground state \cite{friesen_biswas_prl2007,CamposVenuti2007,oh_friesen_prb2010}. We will demonstrate that the functionality of the bus is determined mainly by
the degeneracy of its ground state. For example, a doubly degenerate ground state forms a pseudo-spin, which can be coupled to other spins or
pseudo-spins. A non-degenerate ground state cannot store information, but it may act as a medium for coupling external qubits via virtual
excited states. We will show that an applied magnetic field allows us to navigate between these degenerate and non-degenerate operating modes.
We will further demonstrate that anisotropic interactions may easily be tuned by several different methods.
In addition to varying the magnetic field, these include varying the length of the bus, the average strength of the couplings between the spins, or the connection points between the bus and the external qubits.
Thus, in addition to providing a channel for long range quantum communication, the spin bus also serves as a mediator for anisotropic qubit couplings.
Finally, we will describe an application for anisotropic couplings in a simple, bus-like geometry, in which logical qubits are encoded in double quantum dots in a uniform magnetic field.

\emph{Theoretical model.} Our goal is to compute the effective interactions between a qubit and a spin bus, or between two qubits coupled
through a bus. The bus is composed of two or more spins in a linear geometry, while the qubits may also be composed of one or more spins.  The
key to making this architecture useful for quantum processing is to ensure that the low energy manifold of eigenstates in the bus is separated from all
other energy levels by a gap \cite{friesen_biswas_prl2007}.
Even when the bus is perturbed by the qubit couplings, it
should not be excited outside this manifold.

\begin{figure}
  \includegraphics[width=1.9in]{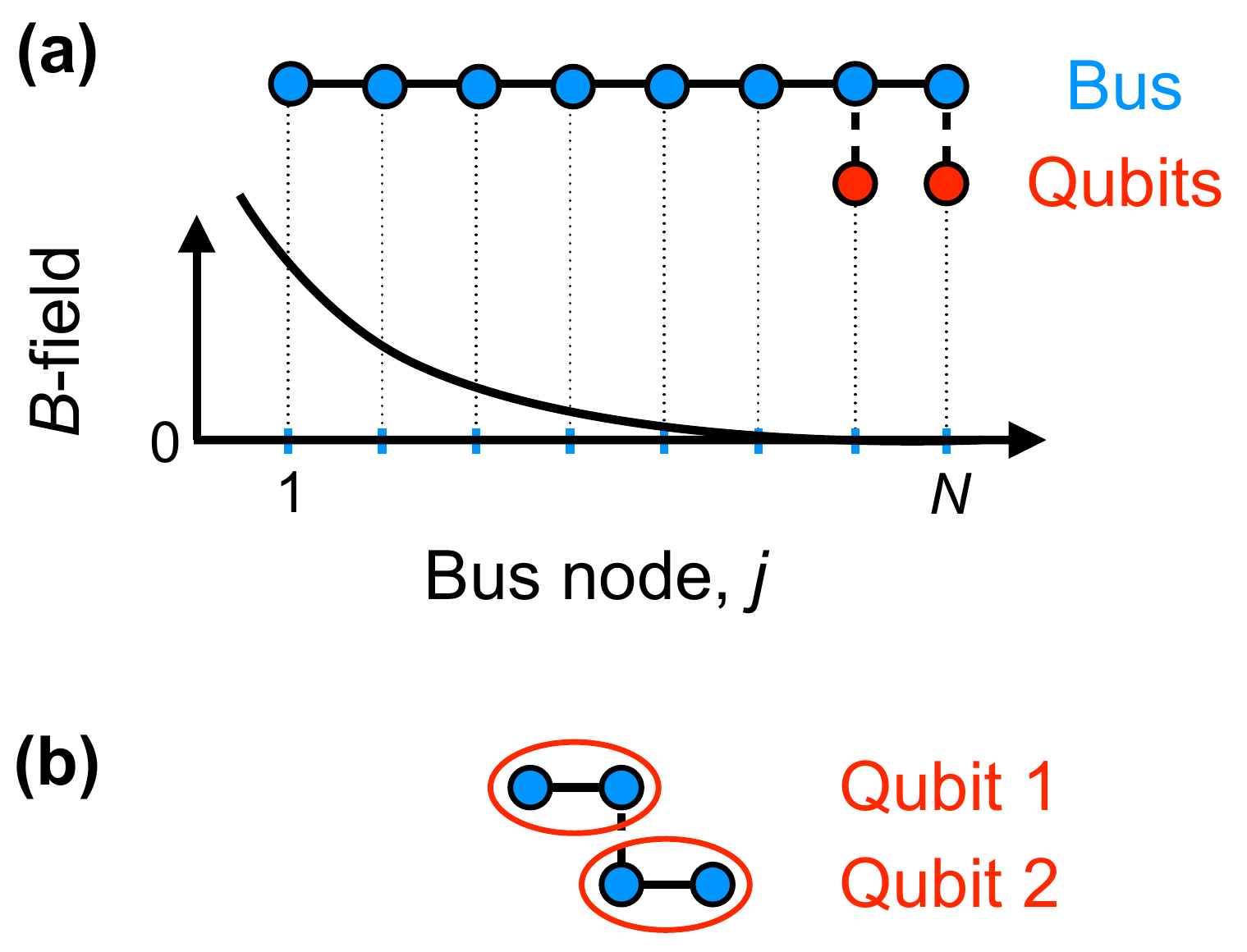}\\
  \caption{(Color online)  Typical architectures for generating XXZ couplings.
           (a) A spatially varying magnetic field applied to a long spin bus, with a small local field on the qubits.
           (b) Coupled two-spin qubits, in a uniform magnetic field.}
  \label{fig:1_model}
\end{figure}

Our method requires the application of a magnetic field.
However, the Zeeman splitting of the qubit states must be small
compared to the energy gap between the ground and excited bus states, to avoid bus excitations.
There are several ways to ensure this.
On possibility is to set up an inhomogeneous magnetic field, as
shown in Fig.~\ref{fig:1_model}(a).  This provides an average, non-zero field on the bus, with a much smaller field on the qubits.  In practice,
large field gradients are difficult to achieve in the laboratory.  However, recent experiments with micromagnets have demonstrated gradients
that are sufficient for our purpose \cite{PioroLadriere2008}.  A long bus can also reduce the gradient requirements and simplify the
experiment.
A second geometry of interest utilizes a small but uniform magnetic field,
which will be easier to implement experimentally.
In this case, we will show that only an even size bus
produces XXZ interactions.  The final geometry we consider overcomes the constraint of small Zeeman splitting.  However, we pay a price that
the external qubit must contain at least two spins.  The simplest geometry of this type is shown in Fig.~\ref{fig:1_model}(b), and will be
discussed at the end of this paper.  In order to provide the most general and least complicated analysis, we now focus on the geometry consisting
of a uniform magnetic field applied to the bus, with a small magnetic field on the qubits.  However, the results we obtain are similar to other
geometries; the emergence of XXZ couplings appears to be ubiquitous.

The full system Hamiltonian is given by
\beq\label{eq:Htot}
H= H_{b} + H_{q} + H_{i} ~,
\eeq
where the unperturbed system consists of bus ($b$) plus qubits ($q$).
The bus consists of an antiferromagnetic spin chain of size $N$, with isotropic bare couplings:
\beq\label{eq:Hb}
H_{b} = J_b \sum_{j=1}^{N-1} \mathbf{s}_{j} \cdot \mathbf{s}_{j+1}
                  - B_b \sum_{j=1}^{N} s_{j,z} ~.
\eeq Here, $J_b$ is the intra-bus coupling strength, assumed to be uniform, $\mathbf{s}_j$ is the spin operator for the $j$th bus node, and
$\mathbf{B}_b$ is the uniform external magnetic field on the bus.  Throughout this paper, we adopt $J_b$ as our energy unit, and we define the
direction of $\mathbf{B}_b$ as $\hat{\mathbf{z}}$.

We allow a qubit to be coupled to any of the bus nodes, with the unperturbed qubit Hamiltonian given by
\beq\label{eq:Hq} H_q = \sum_{j=1}^{N} \mathbf{B}_{j} \cdot
\mathbf{S}_{j}  ~. \eeq
Here, $\mathbf{S}_j$ is the spin operator for the $j$th qubit, and $\mathbf{B}_j$ is its local magnetic field, which can be modified by a local current or nanomagnet \cite{PioroLadriere2008}.
The bus-qubit coupling is assumed to be perturbative, so that it does not
disturb the general manifold structure:
\beq\label{eq:Hi} H_{i} = \sum_{j=1}^{N} J_j \mathbf{S}_{j} \cdot \mathbf{s}_{j}~. \eeq
For qubit $j$,
the coupling $J_j$ may be set to zero; typically, we will only consider one or two qubits
coupled to the bus.
More generally, $J_j$
will be turned on and off as a function of time, in the course of bus operations.

\begin{figure}
  \includegraphics[width=2.9in]{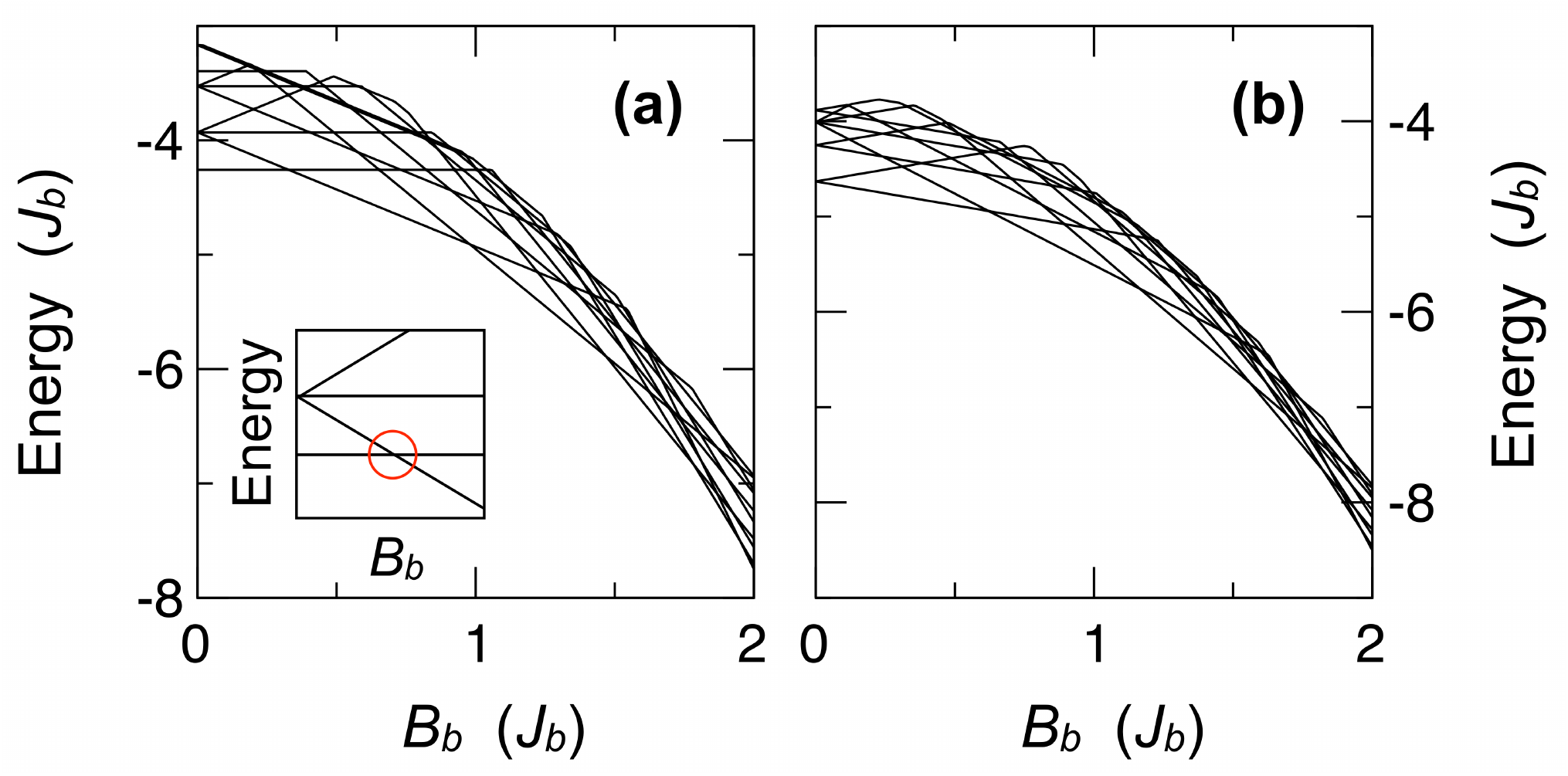}\\
  \caption{(Color online)
    Energy spectra for two different spin buses of size (a) $N=10$ and (b) $N=11$, as a function of uniform magnetic field $B_b$.
           Note that energies and magnetic fields are expressed in unit of the bus spin coupling $J_b$.
           For each plot, we show just the ten lowest energy levels.
           }
  \label{fig:2_energy_concur}
\end{figure}

Some typical energy level diagrams for the bus are shown in Figs.~\ref{fig:2_energy_concur}(a) and (b).  These energy levels were previously
characterized at zero-field, where it was shown that the operation of the bus depends on its size
\cite{friesen_biswas_prl2007,oh_friesen_prb2010}. In the present work, we go on to show that the key bus characteristic, which determines its
behavior at any field, is the 1- or 2-fold degeneracy of its ground state.
We now perform separate analyses of these two regimes.

\emph{Noncritical regime.}
When the magnetic field on the bus is tuned to be away from a level crossing,
the spin bus is characterized by a unique and non-degenerate ground state.
Therefore, when the qubit-bus couplings $J_j$ are weak, and turned on adiabatically, the qubits cannot affect the state of the bus.
When two or more external qubits are simultaneously coupled to the bus, it can mediate an effective interaction between them.
However, this occurs only via virtual excitations of the bus.

We now derive the effective Hamiltonian for the qubits by treating $H_i$ as a perturbation.
{Since the spectral gap of the spin bus is of order $J_b/N$, the expansion parameter is given by $J_j N/ J_b$, which can easily be made small, experimentally.}
The full Hamiltonian is then projected onto a subspace where the bus is in its ground state manifold \cite{SchriefferWolff}, giving
\beqa\label{eq:Heff_nc}
\widetilde{H}
&=& \varepsilon_0 + \sum_{j=1}^{N} \widetilde{\mathbf{B}}_j \cdot \mathbf{S}_{j}  \\
&&+ \sum_{j>j'}^N \left[ \tilde{J}_{j,j'} \left( S_{j,x} S_{j'\!,x} + S_{j,y} S_{j'\!,y}\right)
                     + \widetilde{\Delta}_{j,j'} S_{j,z} S_{j'\!,z} \right]~, \nonumber
\eeqa where $\varepsilon_m$ is the energy of bus eigenstate $|m\rangle$, and $m=0$ corresponds to the ground state.  Note that $\varepsilon_m$,
and the composition of $|m\rangle$ in the spin basis, are both functions of $B_b$. The effective coupling constants in Eq.~(\ref{eq:Heff_nc})
are given by
\begin{gather}
\widetilde{\mathbf{B}}_j
   = \mathbf{B}_j + J_j \langle 0 | s_{j,z} | 0 \rangle \hat{\mathbf{z}}~, \label{eq:Bq_nc}
\\
\tilde{J}_{j,j'} = -2\sum_{m>0} \frac{J_jJ_{j'}}{\vep_m-\vep_0}
\langle 0 | s_{j,x} | m \rangle \langle m | s_{j'\!,x} | 0 \rangle ~, \label{eq:Jeff_nc}
\\
\widetilde{\Delta}_{j,j'} = -2\sum_{m>0} \frac{J_jJ_{j'}}{\vep_m-\vep_0}
\langle 0 | s_{j,z} | m \rangle \langle m | s_{j'\!,z} | 0 \rangle ~.\label{eq:Jeffz_nc}
\end{gather}
We note that qubit $j$ experiences an effective field $\widetilde{\mathbf{B}}_{j}$ that is modified, at first order in the perturbation, by a
`local field' $\langle 0| s_{j,z} |0\rangle$ at position $j$.
Since $\langle 0| s_{j,z} |0\rangle$
alternates in sign \cite{friesen_biswas_prl2007,oh_friesen_prb2010} and has an inhomogeneous magnitude, this produces a built-in field gradient.  Such gradients can be used
to induce qubit rotations \cite{Petta2005}. The effective qubit couplings $\tilde{J}$ and $\widetilde{\Delta}$ arise at second order, for
reasons described above. Note that when multiple qubits are coupled to the bus simultaneously, the resulting network is fully connected.

\begin{figure}
  \includegraphics[width=2.8in]{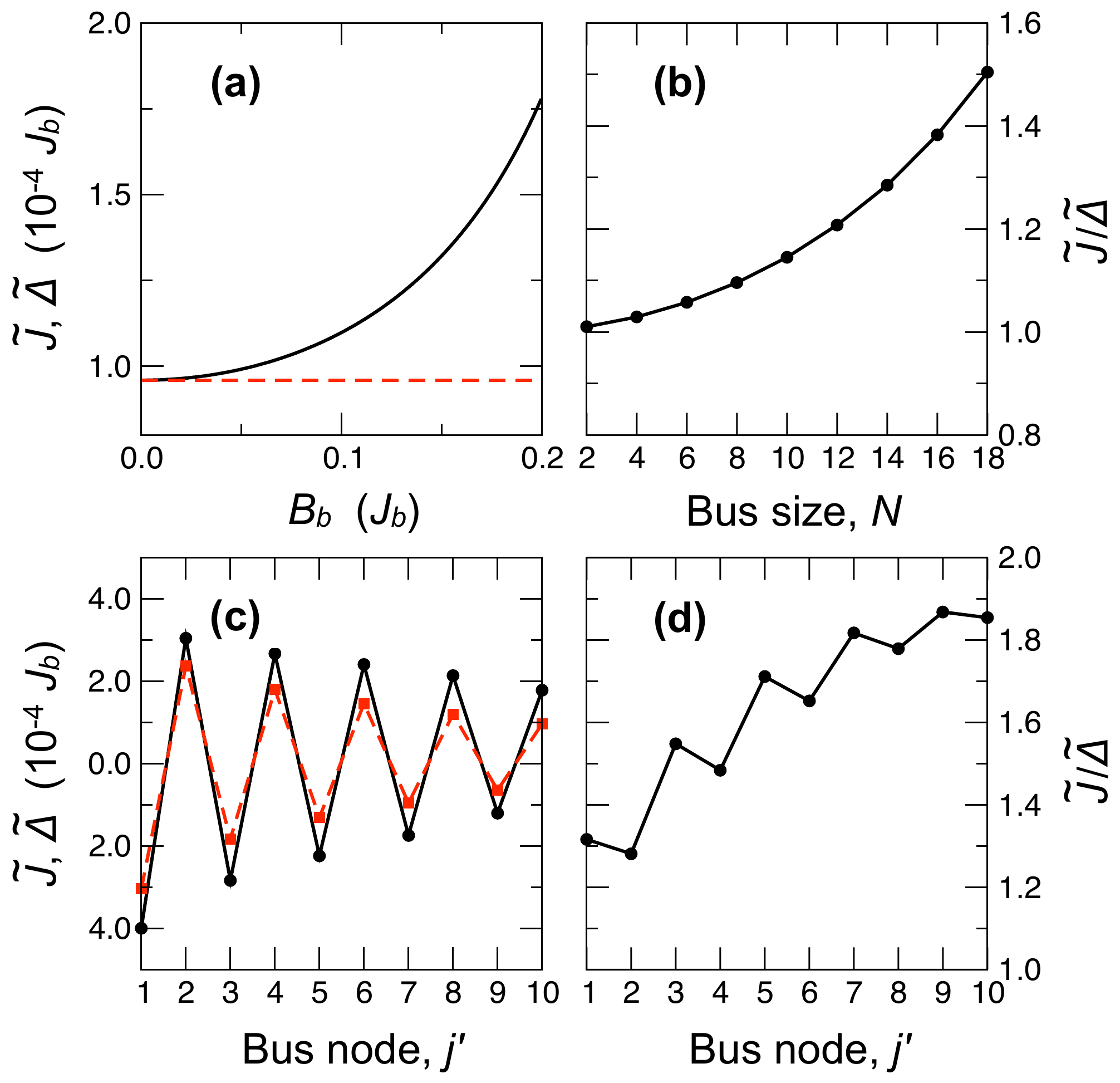}\\
  \caption{(Color online)
           Plots of effective coupling parameters $\tilde{J}$ (solid black) and $\widetilde{\Delta}$ (dashed red), or the coupling
           anisotropy ratio $\tilde{J}/\widetilde{\Delta}$, between two qubits coupled through an even-size bus, with $J_j=0.02\, J_b$
           and a uniform magnetic field.
           (a) The two qubits are coupled to opposite ends of a bus of size $N=10$, with a variable field.
           (b) The two qubits are coupled to opposite ends of a bus with constant field $B_b=0.1\,J_b$ and a variable bus size.
           (c) and (d).  Qubit 1 is attached to node $j=1$ on a bus of size $N=10$, with $B_b=0.2\,J_b$.  Qubit 2 is attached to node $j'$.
           Here, $j'=1$ refers to the case where both qubits are attached to node~1.}
  \label{fig:3_noncritical}
\end{figure}

At zero magnetic field, there is no broken symmetry, so according to Eqs.~(\ref{eq:Jeff_nc}) and (\ref{eq:Jeffz_nc}), we should have
$\tilde{J}=\widetilde{\Delta}$.
At non-zero fields, however, the coupling anisotropy is ubiquitous, even when the field is uniform. This is
demonstrated in Fig.~\ref{fig:3_noncritical}(a), where we take $B_j=B_b\ll J_b$, and consider an even-size bus. The anisotropy
$\tilde{J}/\widetilde{\Delta}$ exhibits an approximate quadratic dependence in both the magnetic field and the bus size.

The preceding example provides an opportunity to explain the emergence of anisotropy.
The effective qubit couplings $\tilde{J}$ and $\widetilde{\Delta}$ are mediated by virtual excitations of the bus from its ground state, which is non-degenerate and spin-0.
The predominant processes involve the low energy, spin-1 bus manifold, which has three eigenstates.
The $x$, $y$ and $z$ components of the effective coupling depend on different excited states, as per Eqs.~(\ref{eq:Jeff_nc}) and (\ref{eq:Jeffz_nc}).
A non-zero magnetic field lifts the degeneracy of the spin-1 manifold, favoring the excitation to the lowest of these energy levels.
This leads to an approximate expression for the anisotropy, given by
$\tilde{J}/\widetilde{\Delta} \approx 1+B_b^2/E_g^2$ where $E_g=\vep_1-\vep_0$ at zero field.
Since the ground state gap $E_g$ depends on bus size as $N^{-1}$
\cite{friesen_biswas_prl2007}, this also explains the quadratic behavior in Fig. \ref{fig:3_noncritical}(b).
{Note that the presence of a spectral gap is essential for spin bus operation.
For both isotropic and anisotropic bus couplings, the gap scales as $N^{-1}$, placing an effective bound on the bus length \cite{friesen_biswas_prl2007}.}
The effective couplings in Eqs.~(\ref{eq:Jeff_nc}) and (\ref{eq:Jeffz_nc}) depend on the local moments of the spin chain.
Thus, uniform and non-uniform magnetic fields will both lift the degeneracy in similar ways.
As an example, the field fluctuations due to background nuclear spins will generate small, inhomogeneous coupling anisotropies of the XYZ type.
We note that the coupling constants alternate in sign, as
shown in panel~(c). The overall magnitude of the coupling decreases slowly with qubit separation,
while the anisotropy generally increases, as shown in panel~(d).

{\it Critical regime.} When the magnetic field on the bus is tuned to be near a ground state level crossing,
the qubits can interact with the bus and affect its state. We characterize the two bus states in the ground state manifold
as pseudo-spins $\ru$ and $\rd$. For definiteness, let us define $\rd$ to be the ground state on the left-hand-side of the level crossing,
and the excited state on the right-hand-side. As before, we derive an effective Hamiltonian,
which now involves the pseudo-spins:
\beqa\label{eq:Heff_c}
&&\widetilde{H} = \vep_\Uparrow \ru \lu
+\vep_\Downarrow \rd \ld
   +\sum_{j=1}^{N} \widetilde{\mathbf{B}}_j \cdot \mathbf{S}_j \\
&& \quad +\sum_{j=1}^{N} \left[ \tilde{J}_{j} \left( S_{j,x} S_{b,x} + S_{j,y} S_{b,y} \right)
                           + \widetilde{\Delta}_j S_{j,z} S_{b,z} \right]~,  \nonumber
\eeqa
with the effective coupling parameters
\begin{gather}
\widetilde{\mathbf{B}}_j = \mathbf{B}_j
    + \frac{J_j}{2} \left( \lu s_{j,z} \ru
                    + \ld s_{j,z} \rd \right) \hat{\mathbf{z}}~, \label{eq:Bj} \\
\tilde{J}_j =   J_j \lu s_{j,+} \rd ~, \label{eq:Jj} \\
\widetilde{\Delta}_j = J_j \left( \lu s_{j,z} \ru
                                    - \ld | s_{j,z} \rd  \right)~. \label{eq:Dj}
\end{gather}
In this case, the coupling anisotropy arises naturally, since the pseudo-spins $\ru$ and $\rd$ generally have different quantum numbers.
The one exception is the odd-size bus near zero-field, for which the pseudo-spins are both spin-1/2.
Note that in the critical regime,
the qubits interact directly with the bus, so the leading
dependence is first order in the perturbation. As a result, the qubit-bus coupling is the same magnitude as the bare exchange coupling.

{\it Pseudo-spin qubits.}
As an application of these results, we now study the two-qubit geometry proposed in Fig.~\ref{fig:1_model}(b).
This scheme is of particular interest, experimentally, due to its small size and the fact that it does not require a large, inhomogeneous magnetic field.
The logical qubits are each composed of two spins with constant coupling $J_b$.
The coupling between the qubits $J_q$ is tunable, and the magnetic field is uniform and tuned
to the crossing point shown in the inset of Fig. \ref{fig:2_energy_concur}(a).
The proposed qubit is a pseudo-spin, similar to the singlet-triplet qubit of Ref.~\cite{Petta2005}.
Solving the problem of coupled pseudo-spins perturbatively,
we obtain the effective interaction between the logical qubits, which takes the same form as Eq.~(\ref{eq:Heff_c}).
The coupling constants are given by $\widetilde{B}=\tilde{J}/4=\widetilde{\Delta}/2=J_q/8$,
corresponding to an anisotropy of 2.
It is interesting to contrast our effective XXZ coupling with the proposed Ising coupling between singlet-triplet qubits \cite{taylor_engel_nphys2005}.
In the latter case, the anisotropy is more extreme, although the overall coupling strength is weaker, since it arises from the electric dipole interaction.
The XXZ scheme is more stable against charge dephasing, and
the two approaches may be viewed as complementary.

In summary, we have presented a scheme to generate anisotropic XXZ couplings in a system of quantum dot spin qubits.
The anisotropy is obtained by combining bare exchange interactions, which are fundamentally isotropic, with an applied field.
By varying the field, we can control the anisotropy while tuning the system through a series of quantum phase transitions.
{Although we specifically considered uniform $J_b$ bus couplings here, we have also investigated the effect of disordered couplings, numerically.
Our main results are essentially unaffected; the quadratic dependence of anisotropy on gap size and magnetic field still holds.
The primary effect of disorder is on the spectral gap $E_g$ \cite{ohpreprint}.
We have previously noted that a large spectral gap will suppress bus excitations and related decoherence mechanisms \cite{friesen_biswas_prl2007}.
We further emphasize that the bus only needs to remain coherent during gate operations, which are fast.}

Our method shows particular promise for pseudo-spin qubits formed
of pairs of spins. To demonstrate the viability of this scheme,
we conclude by estimating the experimental parameters that would be required for
successful operation.  We assume an electron temperature of 100~mK.
In order for the system to equilibrate properly into its ground state
manifold, we require a ground state gap on the order of $J_b=20$~$\mu$eV.
The desired level crossing would occur at a field of 0.2~T, assuming a $g$-factor of 2 for silicon.
$J_q=2$~$\mu$eV provides a perturbative bare qubit coupling, resulting in an effective coupling of
$\tilde{J}=1$~$\mu$eV and $\widetilde{\Delta}=0.5$~$\mu$eV. Device parameters in this range are generally consistent with the current
state-of-the-art. The most challenging aspect in this proposal is the intra-qubit coupling $J_b=20$~$\mu$eV. However, devices with couplings on
the order of 2~$\mu$eV have been demonstrated in GaAs \cite{Petta2005}, and the exchange coupling is known to depend exponentially on
quantum dot separation, so 20~$\mu$eV appears feasible.

This work is supported by the DARPA/MTO QuEST program through a grant from AFOSR and by NSA/LPS through a grant from ARO.
The views and conclusions contained in this document are those of the authors and should not be interpreted as representing official policies, either expressly or implied, of the US Government.

\end{document}